\documentclass[12pt,preprint]{aastex}

\pdfoutput=1

\shorttitle{THREE HYBRID BLAZARS}
\shortauthors{Stanley et al.}

\begin{document}
\title{A MULTIWAVELENGTH STUDY OF THREE HYBRID BLAZARS}

\author{E. C. Stanley\altaffilmark{1}, P. Kharb\altaffilmark{2}, M. L. Lister\altaffilmark{1}, H. L. Marshall\altaffilmark{3}, C. O'Dea\altaffilmark{4,5}, S. Baum\altaffilmark{4,5}}

\altaffiltext{1}{Department of Physics and Astronomy, Purdue University, 525 Northwestern Avenue, West Lafayette, IN 47907, USA}
\altaffiltext{2}{Indian Institute of Astrophysics, II Block, Koramangala, Bangalore 560034, India}
\altaffiltext{3}{Center for Space Research, Room NE80-6031, Massachusetts Institute of Technology, Cambridge, MA 02139, USA}
\altaffiltext{4}{Department of Physics and
Astronomy, University of Manitoba, Winnipeg, MB R3T 2N2 Canada}
\altaffiltext{5}{Rochester Institute of Technology, 84 Lomb Memorial Drive, Rochester, NY 14623, USA}

\begin{abstract}
We present multiwavelength imaging observations of PKS 1045$-$188, 8C 1849+670, and PKS 2216$-$038, three radio-loud active galactic nuclei from the MOJAVE-Chandra Sample that straddle the Fanaroff-Riley (FR) boundary between low- and high-power jets. These hybrid sources provide an excellent opportunity to study jet emission mechanisms and the influence of the external environment. We used archival VLA observations, and new Hubble and Chandra observations to identify and study the spectral properties of five knots in PKS 1045$-$188, two knots in 8C 1849+670, and three knots in PKS 2216$-$038. For the seven X-ray visible knots, we constructed and fit the broadband spectra using synchrotron and inverse Compton/cosmic microwave background (IC/CMB) emission models. In all cases, we found that the lack of detected optical emission ruled out the X-ray emission from the same electron population that produces radio emission. All three sources have high total extended radio power, similar to that of FR\,II sources. We find this is in good agreement with previously studied hybrid sources, where high-power hybrid sources emit X-rays via IC/CMB and the low-power hybrid sources emit X-rays via synchrotron emission. This supports the idea that it is total radio power rather than FR morphology that determines the X-ray emission mechanism. We found no significant asymmetries in the diffuse X-ray emission surrounding the host galaxies. Sources PKS 1045$-$188 and 8C 1849+670 show significant differences in their radio and X-ray termination points, which may result from the deceleration of highly relativistic bulk motion.
\end{abstract}
\keywords{galaxies: active, galaxies: jets, quasars: general, BL Lacertae objects: general}

\section{Introduction}
Radio-loud active galactic nuclei (AGNs) encompass a wide range of objects that are classified by properties such as their luminosity, jet morphology, and orientation with respect to the observer. AGNs with jets that are closely-aligned to our line of sight are called blazars, and their alignment results in apparent superluminal motion, relativistic Doppler boosting, rapid core variability, and high polarization \citep{Angel80,Blandford79}. Blazars comprise the weak optical emission-lined BL Lacertae objects (BL Lacs) and strong emission-lined Flat Spectrum Radio Quasars (FSRQs). The Fanaroff-Riley type I and II \citep[FR\,I, FR\,II;][]{Fanaroff74} classification subdivides radio-loud AGNs based on their luminosity and jet morphology: FR I sources have lower luminosities and jets with their brightest radio-lobe emission located closer to the AGN core than to the jet termination, and FR II sources have higher luminosities and jets with more-distant or terminal hotspot emission. In the standard unification scheme, BL Lacs are associated with FR Is and FSRQ with FR IIs \citep{Urry95}. Factors that may contribute to the FR I and II dichotomy include black hole mass, spin, and accretion rate; jet composition; and the external environment \citep[e.g.,][]{Meier99,Baum95,Wold07}.

High-resolution Chandra X-ray Observatory (CXO) and Hubble Space Telescope (HST) observations of FR\,I and FR\,II radio galaxy jets have revealed systematic differences in their broad-band (i.e., radio-optical-X-ray) spectral energy distributions (SEDs). The jets of FR\,I sources display a concave-downward shape in the $\log\,S_\nu - \log\,\nu$ plot, and can be interpreted as synchrotron emission, while the jets of FR\,II sources display convex-shaped SEDs which can be interpreted as inverse-Compton (IC) emission from up scattered cosmic-microwave background (CMB) photons (the IC/CMB model), or as synchrotron emission from a different electron population than the one producing the radio and optical emission \citep{Worrall09}. However, there are major uncertainties associated with the X-ray emission mechanisms. Creating $\sim$100~kpc jets with highly energetic synchrotron-emitting electrons requires {\it in situ} acceleration along the jet because of their short lifetimes (few tens of years), and the IC/CMB model requires large bulk relativistic motions ($\Gamma_{jet}\sim5$) out to the terminal hot spots, which does not appear to be supported by radio observations \citep{Bridle94}.

The Monitoring Of Jets in Active galactic nuclei with VLBA Experiments \citep[MOJAVE;][]{Lister09} program provides an extensive set of radio observations of AGNs in the northern sky. The original sample consisted of all 135 known AGN with J2000 declination $\delta > -20\arcdeg$, galactic latitude $|b| > 2.5\arcdeg$ and VLBA 15 GHz flux density of at least 1.5 Jy (or 2 Jy for sources with $\delta > 0\arcdeg$) at any epoch between 1994 and 2004. The MOJAVE-Chandra sample \citep[MCS;][]{Hogan11,Kharb12a} is a subset that focuses on AGN that were the best candidates for X-ray jet emission. It consists of 27 MOJAVE quasars and FR\,II radio galaxies that have extended kiloparsec (kpc) scale 1.4 GHz flux density of at least 100 mJy and radio structure at least 3$\arcsec$ in extent. The fraction of hybrid sources is significantly higher in MOJAVE \citep[6\% $-$ 8\%;][]{Kharb10} than in the FIRST survey \citep[1\%;][]{Gawronski06} which was selected for total emission at 1.4 GHz.

Hybrid radio morphology sources have an FR\,I-type plume-like jet on one side of the radio core and an FR\,II-type collimated jet with a hot spot on the other. Their total radio powers are intermediate between FR\,Is and FR\,IIs \citep[e.g.,][]{GopalKrishna00}. These sources exhibit jets with different radio powers and morphologies on either side of the accretion disk$-$black hole system, so this class of source has the potential to become the touchstone for ideas put forth to explain the FR dichotomy.

Hybrid sources in themselves preclude the presence of different central engines or jet compositions in the FR\,I/FR\,II jets, and point towards external factors driving the FR divide. It has been suggested that hybrid sources may simply be lower power FR\,II sources located in asymmetric environments \citep[e.g.,][]{Miller09}. This could cause deceleration to occur in the jet encountering the environmental asymmetry and convert it into an FR\,I jet \citep{Gawronski06}. For the handful of hybrid sources that have been observed with Chandra, the X-ray jets/knots in some cases have concave-downward shaped SEDs like FR\,Is \citep[e.g.,][]{Birkinshaw02} and convex-shaped SEDs like FR\,IIs in others \citep[e.g.,][]{Sambruna08}. It is possible that this is a result of small number statistics and/or shallow observations (typical exposures $\le30$ ks). To this end, we have acquired deep Chandra, HST, and VLA observations of three hybrid blazars, {\it viz.,} PKS 1045$-$188, 8C 1849+670 and PKS 2216$-$038, from the MOJAVE-Chandra sample (Table \ref{table:sources}).

\begin{deluxetable}{lcccccc}
\tabletypesize{\small} 
\tablewidth{0pt}  
\tablecaption{Source Properties\label{table:sources}}  
\tablehead{\colhead{Name} & \colhead{RA (J2000)} & \colhead{Dec (J2000)} &
\colhead{Redshift} & \colhead{Scale (kpc/$\arcsec$)} & \colhead{$\beta_{app}$ (c)} & \colhead{$\theta$ ($\arcdeg$)} \\
\colhead{(1)} & \colhead{(2)} & \colhead{(3)} & \colhead{(4)} & \colhead{(5)} & \colhead{(6)} & \colhead{(7)}}
\startdata
PKS 1045$-$188 & $10^h48^m6^s.621$ & $-19^{\circ}9\arcmin35.727\arcsec$ & 0.595 & 6.653 & 10.51 & 10.87 \\
8C 1849$+$670 & $18^h49^m16^s.072$ & $67^{\circ}5\arcmin41.680\arcsec$ & 0.657 & 6.954 & 23.08 & 4.96 \\
PKS 2216$-$038 & $22^h18^m52^s.038$ & $-3^{\circ}35\arcmin36.879\arcsec$ & 0.901 & 7.812 & 6.73 & 16.9 \\
\enddata
\tablecomments{Columns are as follows: (1) Source name (2) Right ascension (J2000) (3) Declination (J2000) (4) Distance scale (5) Maximum apparent parsec-scale jet speed in units of the speed of light \citep{Lister13} (6) Derived maximum parsec-scale viewing angle using $\theta = 2 \arctan(1/\beta_{app})$}
\end{deluxetable} 

In this paper we present the results of these multi-wavelength observations. It is structured as follows: In section 2 we detail our observations and data reduction. In section 3 we describe the reduced images. In section 4 we discuss the SED generation and model behaviour. In section 5 we discuss our results, and in section 6 we summarize our findings. Throughout this paper we have adopted a cosmology with $H_0$ = 71~km~s$^{-1}$~Mpc$^{-1}$, $\Omega_\lambda$ = 0.73, $\Omega_m$ = 0.27.

\section{Observations and Data Analysis}
We carried out Chandra and HST observations of the three blazars in March, July and October of 2013 (Table~\ref{table:observations}). We describe below our multi-wavelength observations made with Chandra, HST and VLA.

Our observations were not simultaneous, especially in the case of the archival VLA data. Although these sources can be classified as blazars, which typically have rapidly-variable cores, we are studying the kpc-scale jet knots which are significantly downstream from the cores. The kpc-scale jet knots are large in extent, and given their observed radio brightness temperature of $\sim10^5$ K likely have light crossing times of at least 5000 years, neglecting possible beaming effects. So far, the only known case where a kpc-scale jet knot shows short-term variability is the knot HST-1 in M87 \citep{Kovalev07}. This should be considered an exceptional case due to M87's proximity (z = 0.0044) and HST-1's proximity to its core ($\sim$ 1\arcsec, or $\sim$ 80 pc projected).

\begin{deluxetable}{lccc}
\tablecolumns{5} 
\tabletypesize{\small} 
\tablewidth{0pt}  
\tablecaption{HST and Chandra Observations \label{table:observations}}  
\tablehead{\colhead{Source Name} & \colhead{Telescope} & \colhead{Obs. Date} & \colhead{Exp. Time (s)}}
\startdata
PKS 1045$-$188 & Hubble  & 2013 May 16 & 2700  \\
PKS 1045$-$188 & Chandra & 2013 Mar 18 & 56000 \\
8C 1849+670   & Hubble  & 2013 Jul 19 & 3000   \\
8C 1849+670   & Chandra & 2013 Sep 02 & 34000  \\
8C 1849+670   & Chandra & 2013 Sep 05 & 46000  \\
PKS 2216$-$038 & Hubble  & 2013 Jul 30 & 2700  \\ 
PKS 2216$-$038 & Chandra & 2013 Jul 30 & 56000 \\
\enddata
\end{deluxetable} 

\subsection{\label{xray}Chandra X-ray Observations}
We carried out the Chandra observations using the AXAF CCD Imaging Spectrometer (ACIS)$-$S3 chip (which is back-illuminated for low-energy response) in the very faint (VFAINT) timed mode. In order to minimise pileup from the bright blazar cores, we used the 1/8 subarray mode (frame time = 0.441 s). We specified a range of roll angles so that the charge transfer trail of the blazar core did not contaminate the jet emission.

In our X-ray observations, only one counter jet was detected and not every approaching jet knot seen in prior radio observations was detected. A summary of jet type and knot detections is given in Table \ref{table:detections}.

\begin{deluxetable}{lcccccccc}
\tablecolumns{8}
\tabletypesize{\small}
\tablewidth{0pt}
\tablecaption{Jet Knot Specific Flux Densities \label{table:detections}}
\tablehead{\colhead{Name} & \colhead{Component} & \colhead{Type} & \colhead{1.4 GHz} & \colhead{4.8 GHz} & \colhead{8.4 GHz} & \colhead{1537 nm} & \colhead{477 nm} & \colhead{1 keV} \\
& & & \colhead{(mJy)} & \colhead{(mJy)} & \colhead{(mJy)} & \colhead{($\mu$Jy)} & \colhead{($\mu$Jy)} & \colhead{(nJy)} 
}
\startdata
PKS 1045$-$188 & Knot A      & FR-II &  26.7 &  8.9 &        4.2 &    $\leq$ 1.9 &   $\leq$ 0.28 &        1.0 \\
               & Knot B      & FR-II &  33.6 & 12.0 &        3.0 &    $\leq$ 9.0 &   $\leq$ 0.55 &        2.0 \\
               & Knot C      & FR-II &  64.7 & 23.9 &       13.6 &         0.137 &         0.152 &        0.9 \\
               & Knot D      & FR-II &  21.2 &  6.6 & $\leq$ 3.3 &  $\leq$ 0.204 &  $\leq$ 0.136 & $\leq$ 0.1 \\
               & Knot E      & FR-II &  25.0 &  6.1 & $\leq$ 3.3 &        0.0825 &         0.284 & $\leq$ 0.1 \\
               & Counter Jet & FR-I  & 190.5 & 68.3 & $\leq$ 3.3 & $\leq$ 0.0087 & $\leq$ 0.0017 &        1.1 \\
8C 1849+670    & Knot A      & FR-I  &   6.1 &  2.5 &        0.6 & $\leq$ 0.0054 & $\leq$ 0.0017 &        2.2 \\
               & Knot B      & FR-I  &   5.6 &  3.3 &        2.2 & $\leq$ 0.0054 & $\leq$ 0.0017 & $\leq$ 0.2 \\
               & Counter Jet & FR-II &  49.1 & 17.1 &       12.4 & $\leq$ 0.0054 & $\leq$ 0.0017 & $\leq$ 0.3 \\
PKS 2216$-$038 & Knot A      & FR-I  &  44.8 & 17.4 &        9.6 & $\leq$ 0.0063 & $\leq$ 0.0016 &        0.9 \\
               & Knot B      & FR-I  & 116.9 & 51.3 &       47.6 & $\leq$ 0.0063 & $\leq$ 0.0016 &        1.3 \\
               & Knot C      & FR-I  &  36.1 & 20.5 &        2.6 & $\leq$ 0.0063 & $\leq$ 0.0016 &        0.5 \\
\enddata
\tablecomments{In all cases, the knots are in the approaching jet. The optical upper limits for knots A, B, and D of PKS 1045$-$188 suffer from significant external contamination (Section \ref{optical}).}
\end{deluxetable}

We extracted X-ray spectra using the Chandra Interactive Analysis of Observations (CIAO) package version 4.6 with calibration database (CALDB) version 4.5.9. We reprocessed the data using the {\tt chandra$\_$repro} script and filtered the data to include only photons with energies in the range $0.5-10$ keV to account for Chandra calibration and low-energy quantum efficiency contamination. The blazar core was bright enough to create a noticeable CCD readout streak. Techniques exist to replace the streak with a background spectrum, but this was not done to avoid biasing it with the background from either side of the source. We used the radio images to determine the sizes and locations of regions for the extraction. For the jet extraction, we used background regions on the same side of the source as the jet but shifted off axis. For the core extraction, we did not use background regions to avoid biasing it with any possible background asymmetry. We then generated source and background spectra and the associated Response Matrix Files (RMFs) and Ancillary Response Files (ARFs) using the {\tt specextract} script. For all jet components, we generated both unbinned spectra and spectra with one count per bin.

We modelled the spectra with an absorbed power law using the XSPEC package. Due to the low number of counts in individual jet knots (i.e. tens), including the hydrogen column density parameter was tested and did not have a significant impact on the fit, so it was frozen at values from the HEASARC calculator. We used a C statistic \citep{Cash79} modified for background subtraction called the W statistic \citep{Wachter79}, which requires the data have at least one count per bin. As verification, we also fit the spectra using the ISIS package with a maximum likelihood statistic and {\tt subplex} fitting method. The ISIS results for both the binned and unbinned data agreed with the XSPEC results for the binned data. The final fit parameters are listed in Table \ref{table:xspectrum}.

\begin{deluxetable}{lcccc}
\tabletypesize{\small}
\tablewidth{0pt}  
\tablecaption{X-ray Spectral Analysis\label{table:xspectrum}}  
\tablehead{\colhead{Name} & \colhead{Component} & \colhead{$N_H$ (10$^{20}$ cm$^{-2}$)} &
\colhead{$\Gamma$} & \colhead{$\nu F_\nu$ ($10^{-15}$ erg s$^{-1}$ cm$^{-1}$)} \\
\colhead{(1)} & \colhead{(2)} & \colhead{(3)} & \colhead{(4)} & \colhead{(5)}}
\startdata
PKS 1045$-$188 & Core & 4.1 & $1.73^{+0.02}_{-0.03}$ & $457.02^{+10.90}_{-10.73}$ \\
               & A    & 4.1 & $1.79^{+0.36}_{-0.31}$ & $2.49^{+0.89}_{-0.72}$ \\
               & B    & 4.1 & $1.85^{+0.26}_{-0.24}$ & $4.89^{+1.21}_{-1.04}$ \\
		       & C    & 4.1 & $1.81^{+0.39}_{-0.33}$ & $2.22^{+0.84}_{-0.67}$ \\
		       & Counter Jet & 4.1 & $1.15^{+0.22}_{-0.17}$ & $2.74^{+0.91}_{-0.74}$ \\
8C 1849+670    & Core & 5.8 & $1.56^{+0.01}_{-0.02}$ & $773.21^{+12.37}_{-12.17}$ \\
  	           & A    & 5.8 & $1.38^{+0.15}_{-0.14}$ & $5.29^{+1.11}_{-0.97}$ \\
PKS 2216$-$038 & Core & 5.8 & $1.68^{+0.03}_{-0.03}$ & $422.79^{+11.87}_{-11.56}$ \\
		       & A    & 5.8 & $1.28^{+0.27}_{-0.24}$ & $2.24^{+0.96}_{-0.74}$ \\
               & B    & 5.8 & $1.44^{+0.26}_{-0.23}$ & $3.09^{+1.11}_{-0.89}$ \\
               & C    & 5.8 & $1.17^{+0.36}_{-0.31}$ & $1.14^{+0.71}_{-0.49}$ \\
\enddata
\tablecomments{Columns are as follows: (1) Source name (2) Region of spectrum extraction (3) Hydrogen column density (4) Photon index of the power law distribution (5) $\nu F_\nu$ at 1 keV. The hydrogen column density was fixed at galactic values, and the 90\% confidence values are given for the other parameters.}
\end{deluxetable} 

In order to study the galactic environment, we looked for asymmetries in the diffuse X-ray emission around each source. We used SAOImage {\tt DS9}, excluded regions of known radio jet emission, and compared the background X-ray counts on each side of the source using half-circular regions centered on each jet. Based on the radio emission, we considered emission within a radius of 23\arcsec ~for PKS 1045$-$188 and 8C 1849+670 and 33\arcsec ~for PKS 2216$-$038. We did this comparison on unbinned, unsmoothed 500-10000 eV data files and on data files that had been filtered to energies of 500-1000, 1000-2000, and 2000-10000 eV. We sampled the background throughout the rest of the image using regions of the same size to estimate the mean and standard deviation for such regions. In all cases, the background around the source was elevated above the mean as one would expect near an AGN, but none of the differences between the half-circular regions were greater than 1.5 times the standard deviation. The percentage differences (i.e. for background regions A and B, $(counts_A - counts_B) / counts_B$) that would have been required to be considered significant (greater than 3$\sigma$) are 18\%, 33\%, and 23\% for PKS 1045$-$188, 8C 1849+670, and PKS 2216$-$038 respectively. Attempts have been made to look at nearby galaxy number densities and asymmetric galactic interactions \citep{Kharb14}; however, results have been inconclusive due to lack of redshift information to confirm group or cluster measurement.

\subsection{\label{optical}HST Optical Observations}

We carried out the HST observations using the Wide Field Camera 3 (WFC3) through the wideband F160W and F475W filters (1537 nm and 477 nm central wavelengths, respectively). We loaded pipeline drizzled images into SAOImage {\tt DS9} and extracted background-subtracted counts using the Chandra-Ed Archive Server analysis commands from the {\tt DS9} Virtual Observatory. We then computed $\lambda F_{\lambda}$ values by multiplying the background-subtracted counts (in units of electron s$^{-1}$) by the inverse sensitivity keyword PHOTFLAM (ergs cm$^{-2}$ \AA$^{-1}$ electron$^{-1}$) and the pivot wavelength keyword PHOTPLAM (\AA) from the image headers.

For the sources 8C 1849+670 and PKS 2216$-$038, we detected no significant optical emission other than the jet core. This lack of detection is meaningful for constraining the X-ray emission mechanism, so three times the standard deviation of the background was used as an upper limit for the SED data points.

For PKS 1045$-$188 jet knots A, B, and D, any possible optical emission was contaminated by multiple stars and galaxies in the field of view. Because a perfect subtraction was impossible for such a crowded field and the optical fluxes were vital for ruling out potential emission models, we took a conservative approach and used direct, unmodified measurements as upper limits for the SED data points. For PKS 1045$-$188 knot C, we used a smaller region of 0.6$\arcsec$ radius to reduce contamination because beyond this radius, when looking in directions away from contamination (i.e., away from the galaxy that overlaps the jet between knots B and C), count rates fell back to background levels.

PKS 1045$-$188 knots C and E have possible optical detections. Small regions of emission coincide well with the peaks of the 4.8 GHz radio knots. To estimate the likelihood of this being coincidence, we performed a simple Poisson probability test. We used the {\tt daofind} task in the DAOPHOT \citep{Stetson87} package in IRAF (Image Reduction and Analysis Facility) version 2.16 \footnote{\url{http://iraf.noao.edu/}} to count the number of stars in the field, and then used Poisson statistics to estimate the likelihood of finding a random source with equal or lesser magnitude in the knot regions. We used regions of 0.5\arcsec\, radius based on the Chandra PSF size. The chance probabilities are 4.4\% and 2.1\% for knot C and E respectively, which supports these being jet detections.

\subsection{\label{radio}VLA Radio Observations}

Pipeline calibrated radio data for all three blazars were available at several frequencies and array configurations of the VLA in the NRAO image archive\footnote{\url{https://archive.nrao.edu/archive/archiveimage.html}}. We specifically examined the VLA A-array data at 1.4, 4.8 and 8.4~GHz (Table~\ref{table:vlaobservations}). We downloaded the calibrated ({\it u,v}) datasets and after additional phase self-calibration, created final images convolved with the appropriate beam-sizes to match the Chandra and HST data. Imaging and further self-calibration were carried out using the {\tt Difmap} package \citep{Shepherd97}. We began with a human-guided {\tt CLEAN} deconvolution to ensure that {\tt CLEAN} windows coincided with regions of actual emission, and once the jet knots were identified, further iterations of {\tt CLEAN}-ing and self-calibration were done using a loop until the model converged. To estimate the fluxes of individual features, we replaced the {\tt CLEAN} components in the region of the jet knots with circular Gaussian models using the {\tt Difmap modelfit} task and then used the model error analysis tool in the {\tt Difwrap} package \citep{Lovell00}. In the event of no detection, we took three times the standard deviation of the background as an upper limit.

\begin{deluxetable}{lcccccc}
\tabletypesize{\small}
\tablewidth{0pt}  
\tablecaption{VLA Observations \label{table:vlaobservations}}  
\tablehead{
\colhead{Name} & \colhead{Frequency} &
\colhead{Project} & \colhead{Obs.} & \colhead{Beam} & \colhead{Position Angle} &
\colhead{Image rms} \\
 & \colhead{(GHz)} & \colhead{ID} & \colhead{Date} &
\colhead{($\arcsec \times \arcsec$)} & \colhead{($^{\circ}$)} & \colhead{(mJy beam$^{-1}$)}}
\startdata
PKS 1045$-$188 & 1.49 & AG361 & 1992 Nov 18 & 3.18$\times$1.29 & $-$29.36 & 0.27 \\
               & 4.86 & AB660 & 1992 Dec 14 & 1.13$\times$0.38 & $-$42.38 & 0.33 \\
               & 8.44 & AG361 & 1992 Nov 18 & 0.47$\times$0.16 & 10.27  & 0.17 \\
8C 1849+670    & 1.51 & AL499 & 1999 Aug 27 & 1.43$\times$1.18 & $-$8.53  & 0.17 \\
               & 4.86 & AP250 & 1993 May 02 & 1.54$\times$1.38 & $-$62.17 & 0.09 \\
               & 8.44 & AL401 & 1997 Aug 26 & 3.19$\times$2.42 & 4.42   & 0.11 \\
PKS 2216$-$038 & 1.40 & AL634 & 2004 Nov 21 & 0.41$\times$0.54 & 27.01  & 0.17 \\
               & 4.86 & AK491 & 1999 Aug 05 & 5.02$\times$3.92 & 17.86  & 0.17 \\
               & 8.46 & AR415 & 1999 Jul 31 & 3.82$\times$2.36 & 39.67  & 0.62 \\
\enddata
\end{deluxetable} 

\section{Image Analysis}

\subsection{\label{1045analysis}PKS 1045$-$188}
\begin{figure}
\includegraphics[width=8cm]{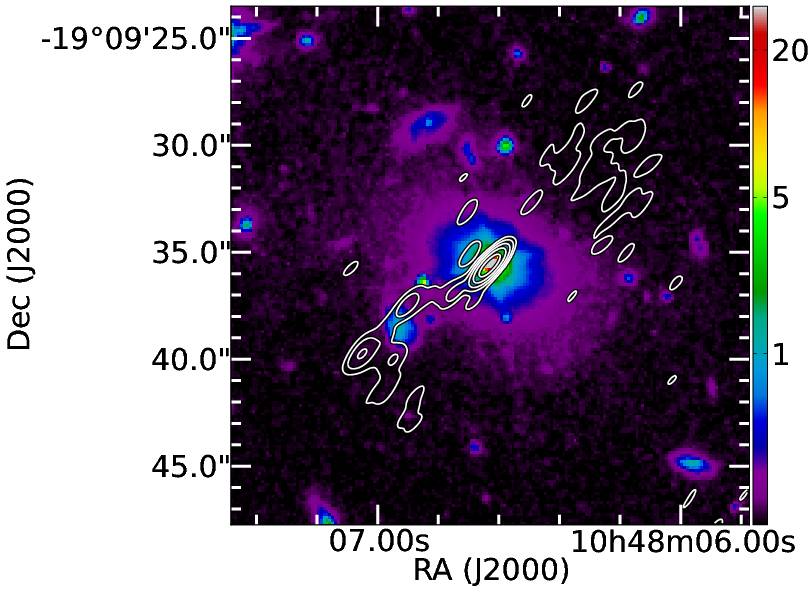}
\includegraphics[width=8cm]{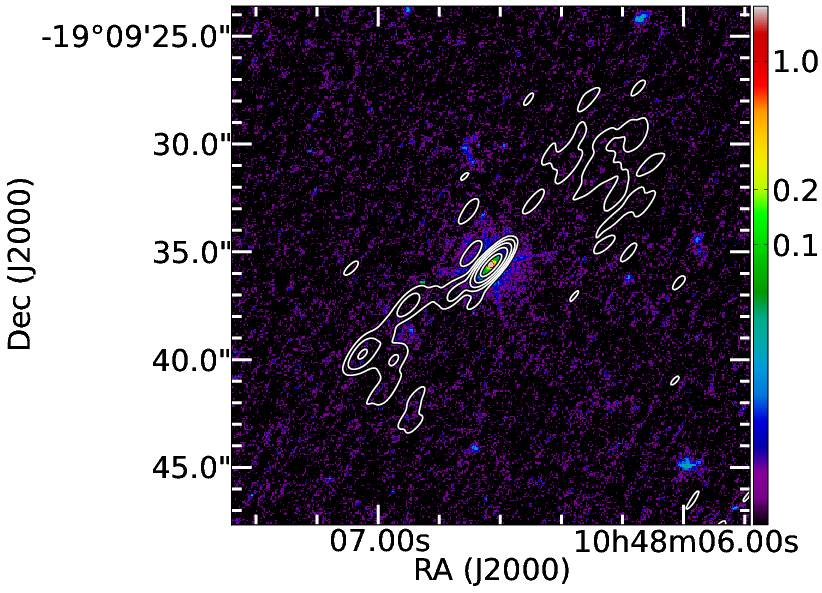}
\caption{VLA 4.8 GHz radio contours superimposed on (left) the colour HST/F160W image and (right) HST/F475W image of PKS 1045$-$188. The color scales correspond to image counts. The lowest contour level is three times the radio image rms, and each higher contour is four times the previous one.}
\label{fig:1045opt}
\end{figure}

\begin{figure}
\includegraphics[width=8cm]{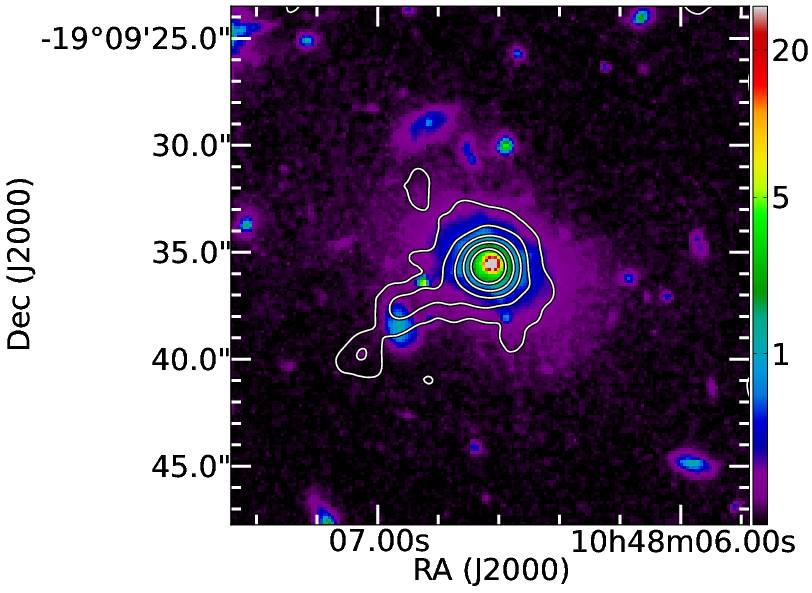}
\includegraphics[width=8cm]{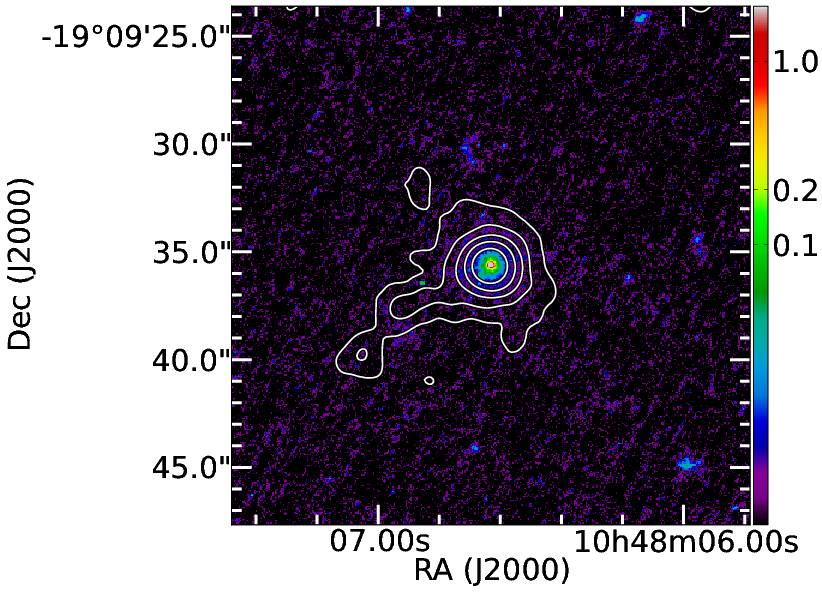}
\caption{Chandra X-ray contours superimposed on (left) the colour HST/F160W image and (right) HST/F475W image of PKS 1045$-$188. The color scales correspond to image counts. The lowest contour level is three times the radio image rms, and each higher contour is four times the previous one.}
\label{fig:1045opt2}
\end{figure}

Given that MOJAVE VLBA imaging shows a one-sided parsec-scale jet at position angle $146\arcdeg$ (measured from north through east), we consider the kiloparsec-scale jet at position angle $125\arcdeg$ \citep{Lister13} to correspond to the approaching jet. The 4.8 GHz VLA image shows five FR\,II approaching jet knots and the FR\,I counter jet plume. Extensive X-ray emission is present along the approaching jet up to and including knot C, after which the radio jet bends $\sim$90$^\circ$ and continues to knots D and E (Figure~\ref{fig:1045rad}). The radio jet bend is likely exaggerated by the small angle between the jet and line of sight ($\theta$), which could well be comparable to the angle of the parsec-scale jet to the line of sight, $\theta \leq 2 \arctan(1/\beta_{max}) = 10.9 \arcdeg$. There is possible X-ray emission coincident with the radio counter jet plume, but it is too faint and diffuse to concretely identify it as counter jet emission without deeper exposure. Any possible optical emission from knots A and B is obscured by or blended with two field galaxies. There is faint optical emission at the sites of knots C and E, but no optical emission is present at the locations of knot D or the counter jet lobe (Figures~\ref{fig:1045opt} and \ref{fig:1045opt2}).

\subsection{8C 1849+670}
\begin{figure}
\includegraphics[width=8cm]{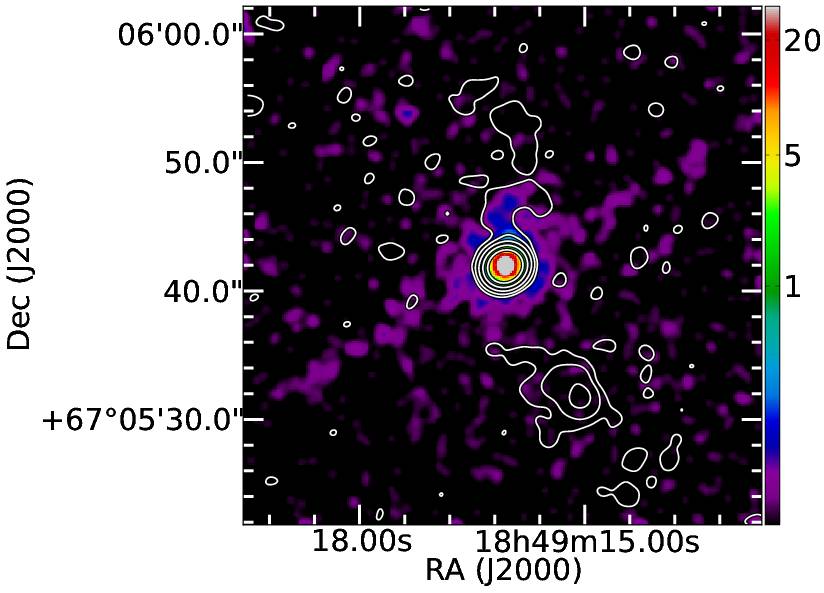}
\includegraphics[width=8cm]{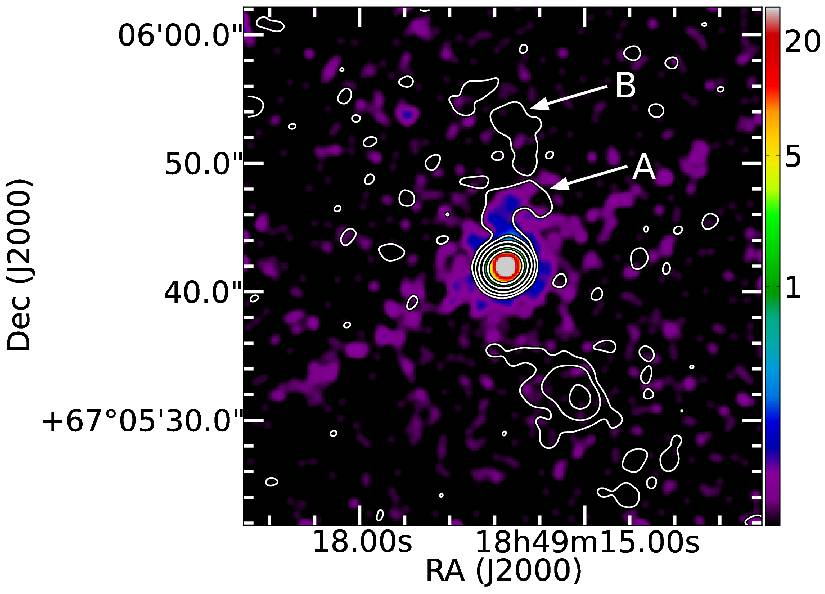}
\caption{(left) Chandra X-ray image of 8C 1849+670 in colour superimposed by VLA 4.8 GHz radio contours. (right) Jet knots identified. The color scales correspond to image counts. The lowest contour level is three times the radio image rms, and each higher contour is four times the previous one.}
\label{fig:1849rad}
\end{figure}

\begin{figure}
\includegraphics[width=8cm]{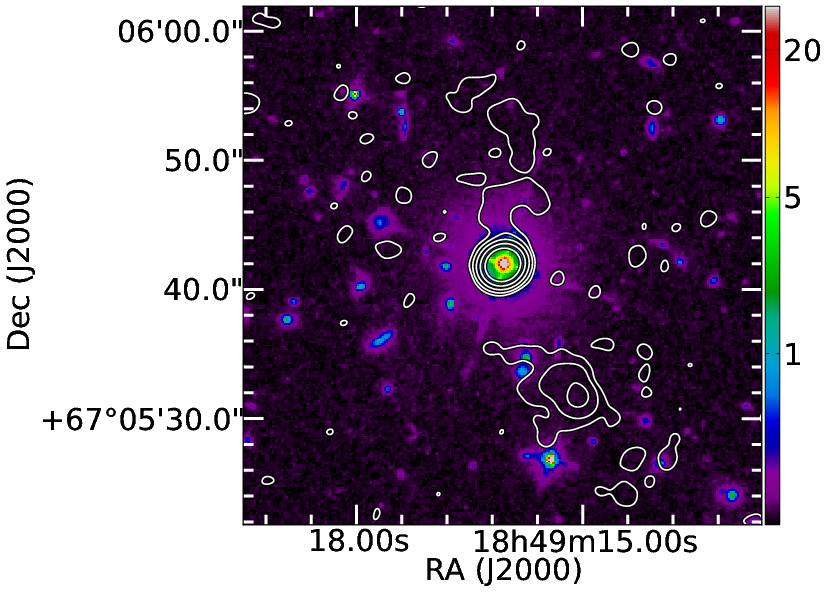}
\includegraphics[width=8cm]{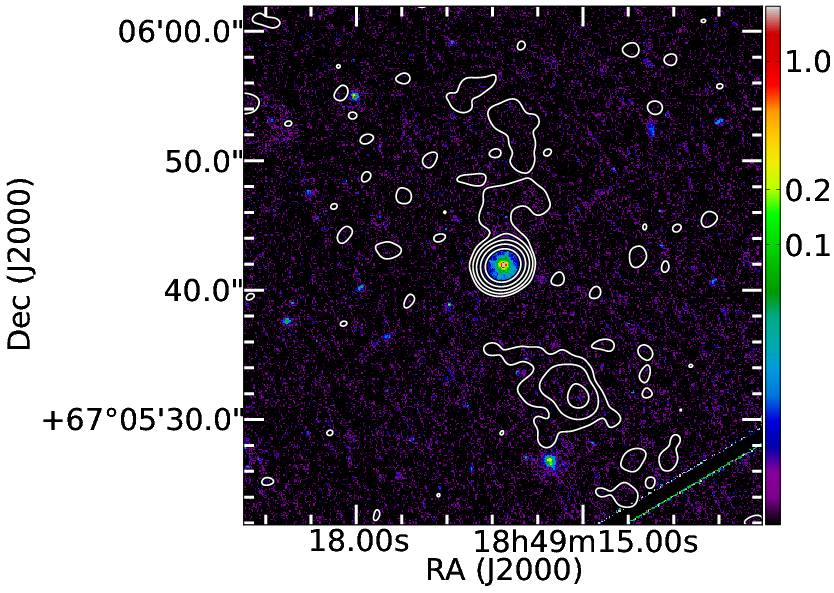}
\caption{VLA 4.8 GHz radio contours superimposed on (left) the colour HST/F160W image and (right) HST/F475W image of 8C 1849+670. The color scales correspond to image counts. The lowest contour level is three times the radio image rms, and each higher contour is four times the previous one.}
\label{fig:1849opt}
\end{figure}

\begin{figure}
\includegraphics[width=8cm]{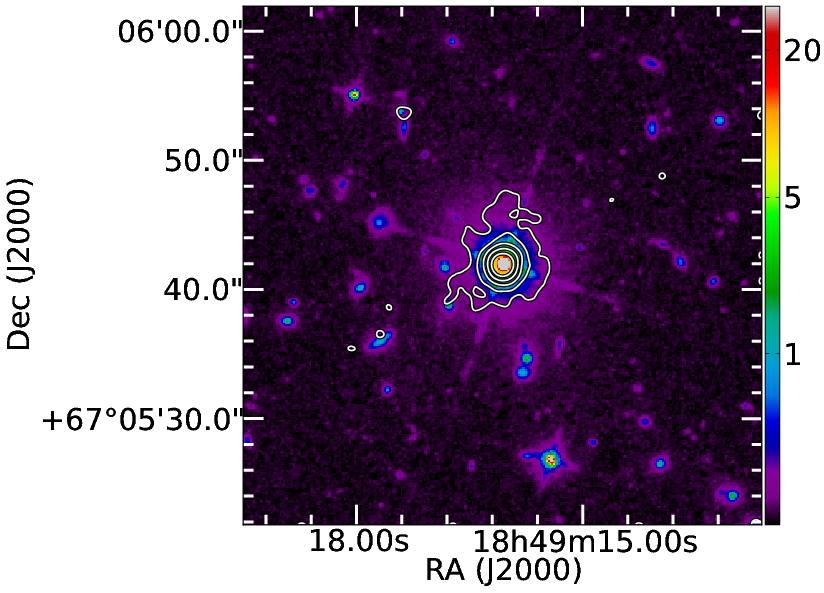}
\includegraphics[width=8cm]{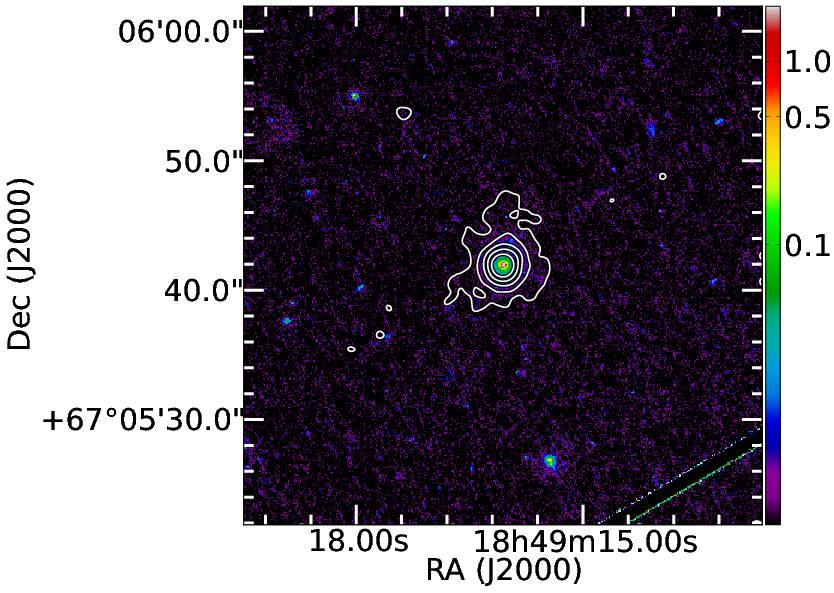}
\caption{Chandra X-ray contours superimposed on (left) the colour HST/F160W image and (right) HST/F475W image of 8C 1849+670. The color scales correspond to image counts. The lowest contour level is three times the radio image rms, and each higher contour is four times the previous one.}
\label{fig:1849opt2}
\end{figure}

Given that MOJAVE VLBA imaging shows a one-sided parsec-scale jet at position angle $-52\arcdeg$, we consider the kiloparsec-scale jet at position angle $0\arcdeg$ \citep{Lister13} to correspond to the approaching jet. The 1.4 GHz VLA image shows two FR\,I approaching jet knots and the terminal FR\,II counter jet hotspot. X-ray emission is detectable only from the approaching jet knot A (Figure~\ref{fig:1849rad}). No optical emission is detected from either jet (Figures~\ref{fig:1849opt} and \ref{fig:1849opt2}).

\subsection{PKS 2216$-$038} 
\begin{figure}
\includegraphics[width=8cm]{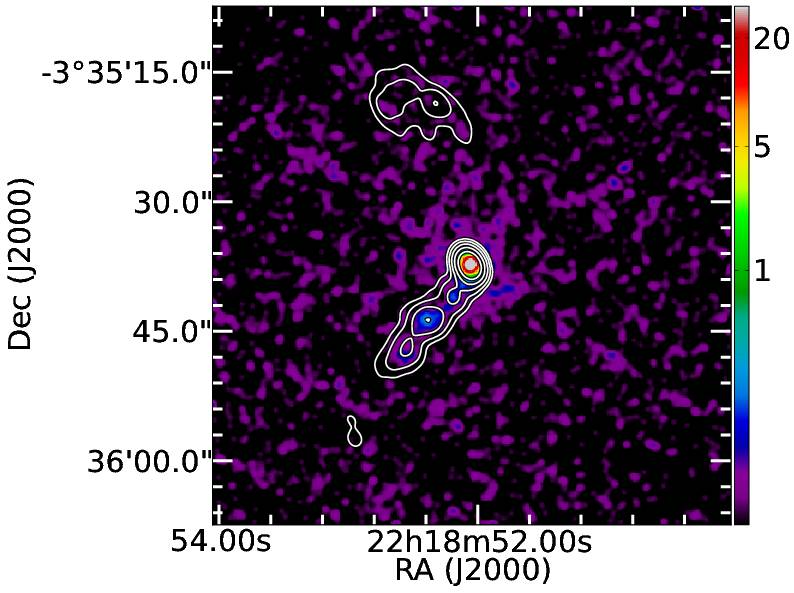}
\includegraphics[width=8cm]{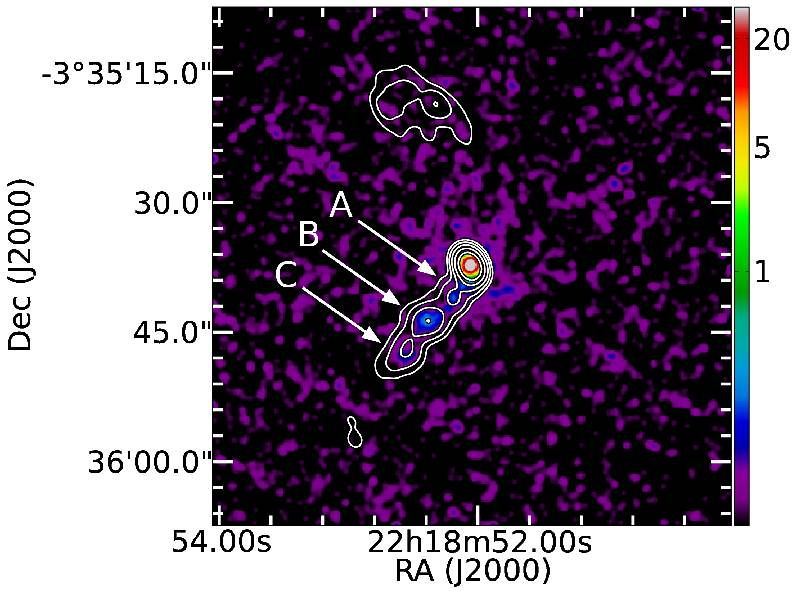}
\caption{(left) Chandra X-ray image of PKS 2216$-$038 in colour superimposed by VLA 1.4 GHz radio contours. (right) Jet knots identified. The color scales correspond to image counts. The lowest contour level is three times the radio image rms, and each higher contour is four times the previous one.}
\label{fig:2216rad}
\end{figure}

\begin{figure}
\includegraphics[width=8cm]{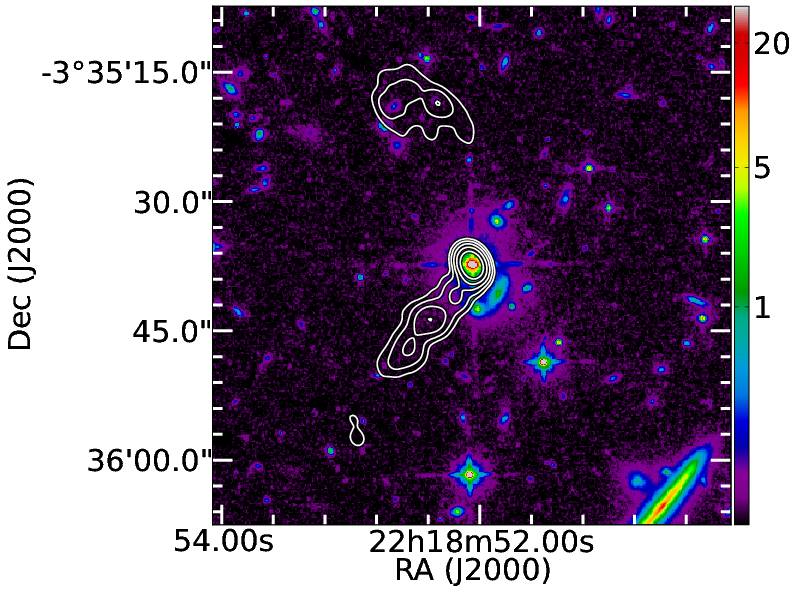}
\includegraphics[width=8cm]{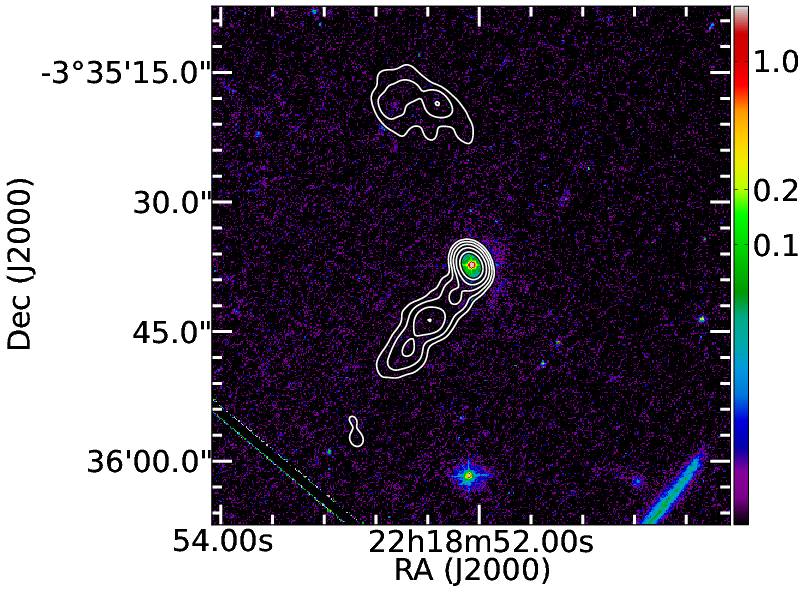}
\caption{VLA 1.4 GHz radio contours superimposed on (left) the colour HST/F160W image and (right) HST/F475W image of PKS 2216$-$038. The color scales correspond to image counts. The lowest contour level is three times the radio image rms, and each higher contour is four times the previous one.}
\label{fig:2216opt}
\end{figure}

\begin{figure}
\includegraphics[width=8cm]{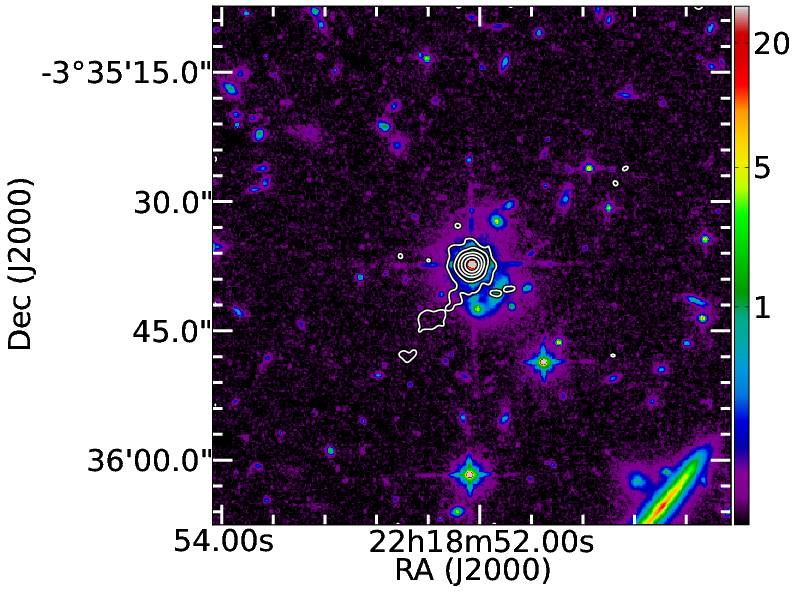}
\includegraphics[width=8cm]{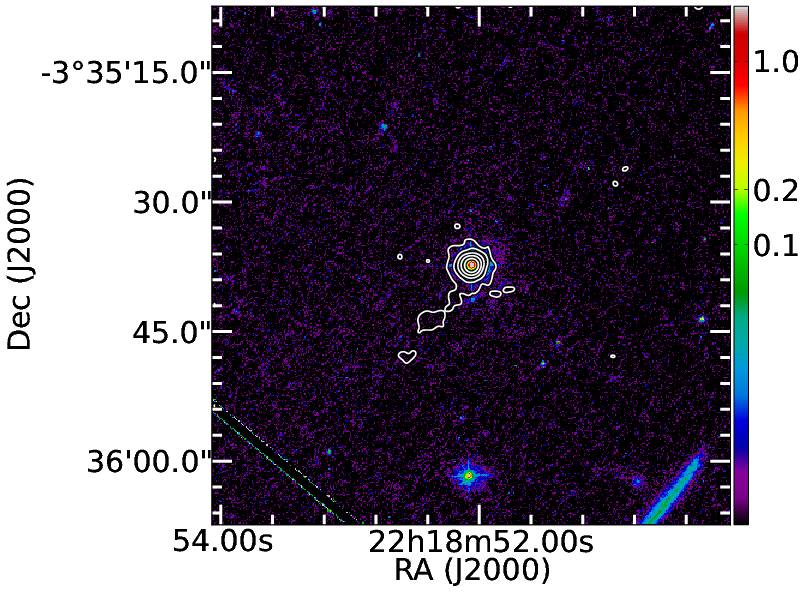}
\caption{Chandra X-ray contours superimposed on (left) the colour HST/F160W image and (right) HST/F475W image of PKS 2216$-$038. The color scales correspond to image counts. The lowest contour level is three times the radio image rms, and each higher contour is four times the previous one.}
\label{fig:2216opt2}
\end{figure}

Given that MOJAVE VLBA imaging shows a one-sided parsec-scale jet at position angle $-170\arcdeg$, we consider the kiloparsec-scale jet at position angle $135\arcdeg$ \citep{Lister13} to correspond to the approaching jet. The 1.4 GHz VLA image shows three FR\,I approaching jet knots and the FR\,II counter jet hotspot. Extensive X-ray emission is present throughout the approaching jet (Figure~\ref{fig:2216rad}). The HST image shows a field galaxy near the optical core position, but no clear jet emission (Figures~\ref{fig:2216opt} and \ref{fig:2216opt2}). It is not clear if this galaxy is physically close to the host galaxy of PKS 2216$-$038 because no spectroscopic information is available on it.

\begin{figure}
\includegraphics[width=8cm]{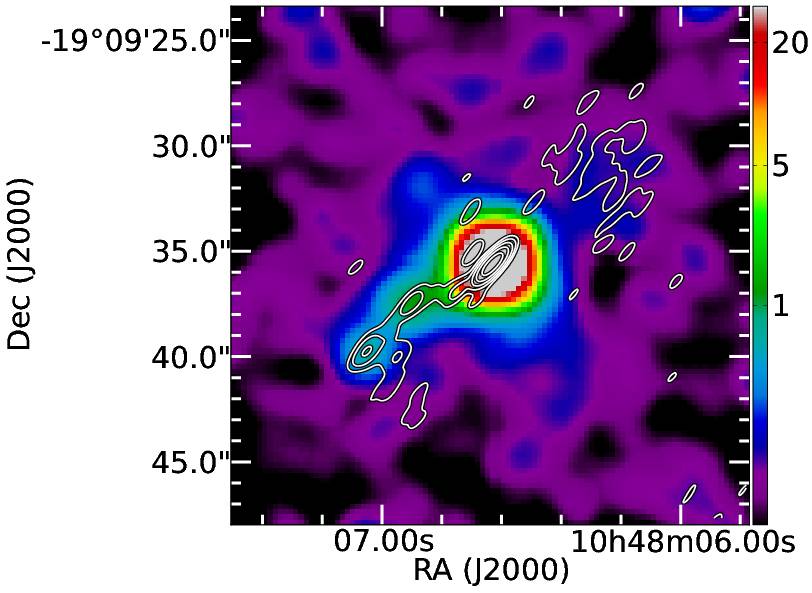}
\includegraphics[width=8cm]{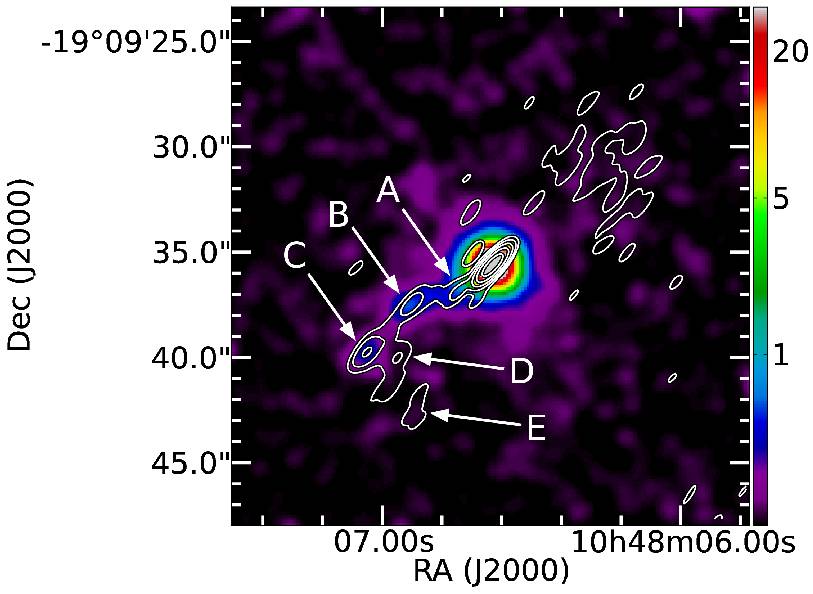}
\caption{(left) Chandra X-ray image of PKS 1045$-$188 in colour superimposed by VLA 4.8 GHz radio contours with contrast adjusted to show the possible counter jet emission (Section \ref{1045analysis}). (right) Jet knots identified. The color scales correspond to image counts. The lowest contour level is three times the radio image rms, and each higher contour is four times the previous one.}
\label{fig:1045rad}
\end{figure}



\section{\label{seds}SED Generation}
\begin{figure}
\centering{
\includegraphics[width=9cm]{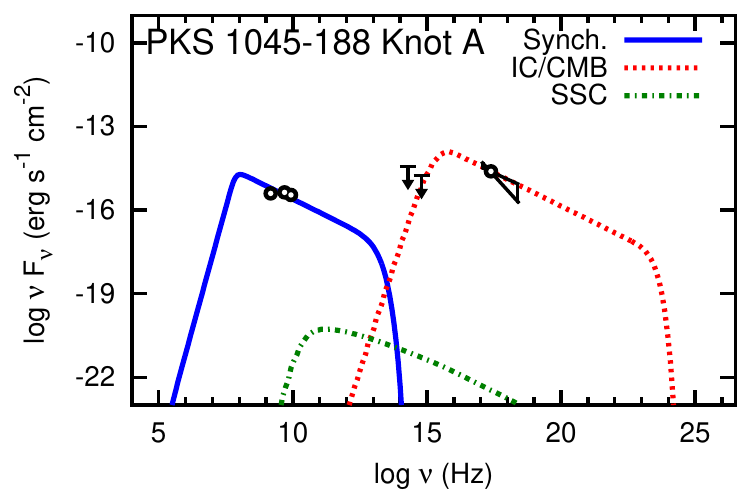}
\includegraphics[width=9cm]{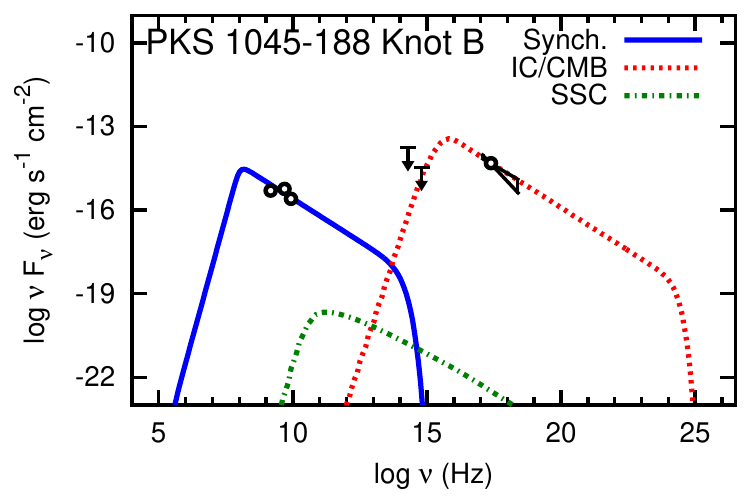}
\includegraphics[width=9cm]{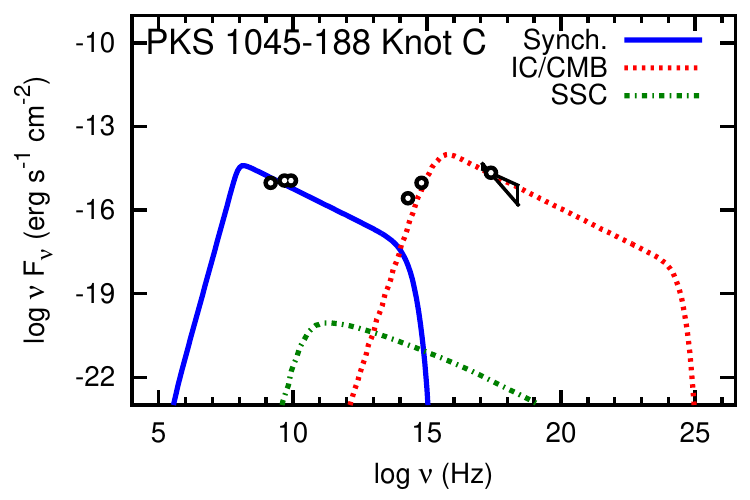}}
\caption{Broad-band spectral energy distributions (SEDs) of (top) knot A, and (middle) knot B, (bottom) knot C of PKS 1045$-$188. The solid blue line, red dotted line, and green-dotted-dashed line are synchrotron, IC/CMB, and SSC spectral components, respectively.}
\label{fig:1045sed}
\end{figure}

\begin{figure}
\centering{
\includegraphics[width=9cm]{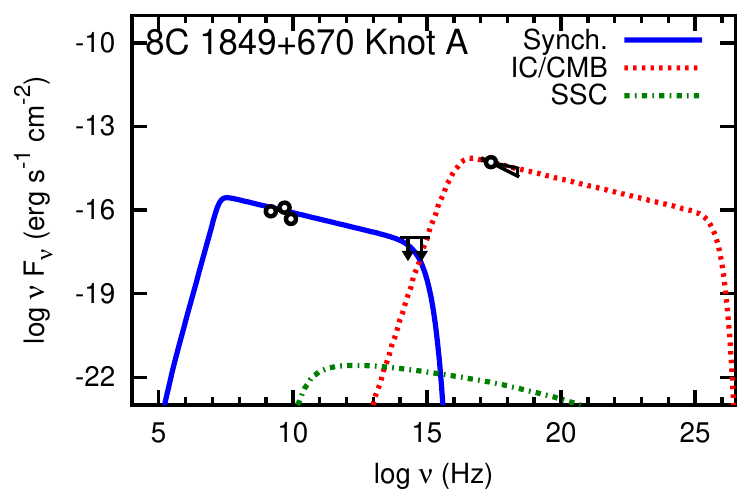}}
\caption{Broad-band SEDs of knot A in the 8C 1849+670 jet. The solid blue line, red dotted line, and green-dotted-dashed line are synchrotron, IC/CMB, and SSC spectral components, respectively.}
\label{fig:1849sed}
\end{figure}

\begin{figure}
\centering{
\includegraphics[width=9cm]{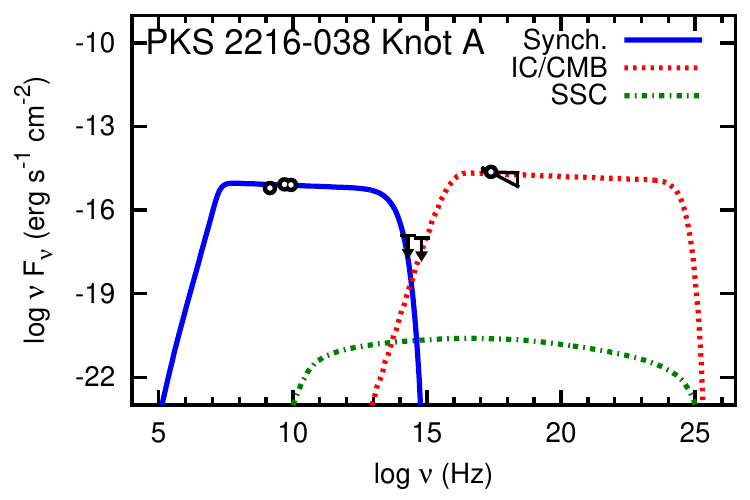}
\includegraphics[width=9cm]{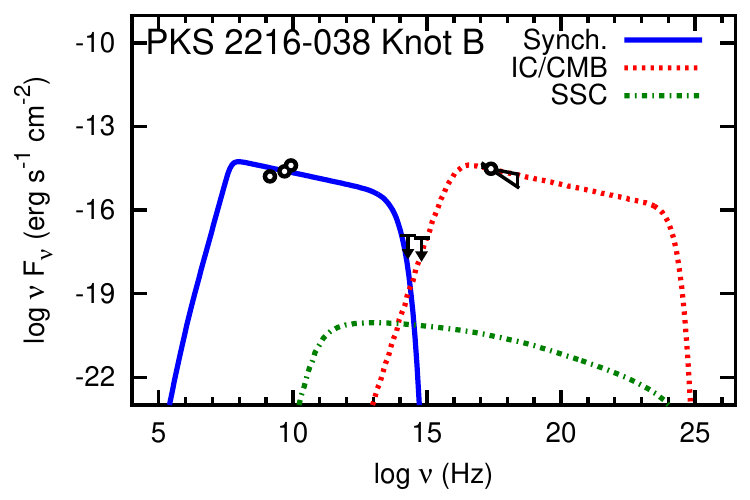}
\includegraphics[width=9cm]{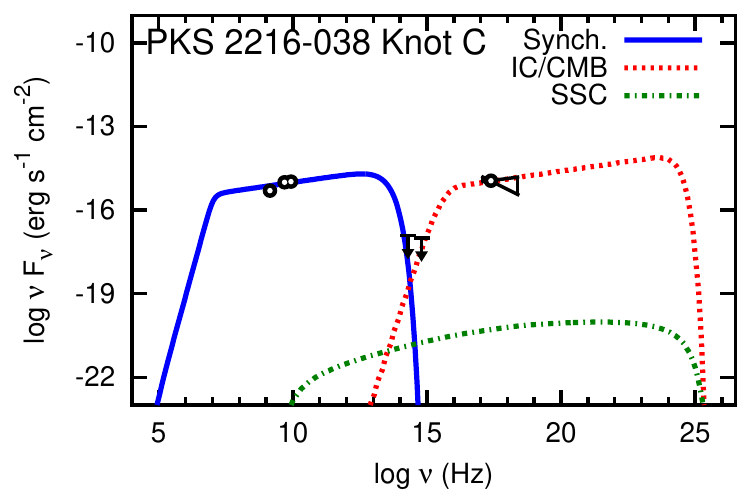}}
\caption{Broad-band SEDs of knots A, B and C in the PKS 2216$-$038 jet. The solid blue line, red dotted line, and green-dotted-dashed line are synchrotron, IC/CMB, and SSC spectral components, respectively.}
\label{fig:2216sed}
\end{figure}

We used the jet emission model of \citet{Krawczynski04} to construct fits to the jet knot SEDs. It generates spectral energy distributions for synchrotron, inverse-Compton (IC) of cosmic microwave background (CMB) photons, and synchrotron self-Compton (SSC) emission for a given set of physical jet parameters. We used magnetic field and bulk flow Doppler factor values calculated from VLBI observations by \citet{Hogan11} as a starting point and adjusted them to fit. The SEDs are presented in Figures \ref{fig:1045sed}, \ref{fig:1849sed}, \ref{fig:2216sed} and the SED best-fit parameters are given in Table \ref{table:sed}.

\begin{deluxetable}{lccccccc}
\tablewidth{0pt}  
\tablecaption{SED best-fit parameters\label{table:sed}}  
\tablehead{\colhead{Name} & \colhead{Knot} & \colhead{$\delta$} & \colhead{$B$} & \colhead{$w_{p_{soll}}$} & \colhead{$\gamma_{min}$} & \colhead{$\gamma_{max}$} & \colhead{$n$} \\
& & & \colhead{($\mu$G)} & \colhead{($10^{-12}$ erg cm$^{-3}$)} & & \colhead{($10^{5}$)} & \\
\colhead{(1)} & \colhead{(2)} & \colhead{(3)} & \colhead{(4)} & \colhead{(5)} &\colhead{(6)} & \colhead{(7)} & \colhead{(8)}}
\startdata
PKS 1045$-$188 & A & 4.1 $\pm$ 0.7 & 45 $\pm$ 6  & 5 $\pm$ 12   & 12 $\pm$ 17 & 1 $\pm$ 9    & 4.0 $\pm$ 0.1 \\
               & B & 4.1 $\pm$ 0.7 & 50 $\pm$ 8  & 13 $\pm$ 29  & 14 $\pm$ 19 & 2.5 $\pm$ 9  & 4.2 $\pm$ 0.2 \\
               & C & 4.1 $\pm$ 0.7 & 70 $\pm$ 10 & 4 $\pm$ 13   & 12 $\pm$ 2  & 2.5 $\pm$ 14 & 4.0 $\pm$ 0.1 \\
1849$+$670     & A & 4.2 $\pm$ 0.7 & 10 $\pm$ 4  & 0.35 $\pm$ 2 & 30 $\pm$ 14 & 12 $\pm$ 10  & 3.5 $\pm$ 0.2 \\
PKS 2216$-$038 & A & 3.0 $\pm$ 0.9 & 20 $\pm$ 7  & 0.19 $\pm$ 3 & 26 $\pm$ 16 & 3.5 $\pm$ 2  & 3.1 $\pm$ 0.3 \\
               & B & 3.0 $\pm$ 0.8 & 50 $\pm$ 3  & 0.29 $\pm$ 3 & 32 $\pm$ 17 & 2 $\pm$ 2    & 3.4 $\pm$ 0.1 \\
               & C & 3.0 $\pm$ 0.9 & 15 $\pm$ 4  & 0.1 $\pm$ 1  & 20 $\pm$ 19 & 3.3 $\pm$ 2  & 2.7 $\pm$ 0.3 \\
\enddata
\tablecomments{Columns are as follows: (1) Source name (2) Region of spectrum extraction and fitting (3) Bulk flow Doppler factor (4) Magnetic field strength (5) Electron energy density (6) Minimum electron Lorentz factor (7) Maximum electron Lorentz factor (8) Power-law index of the electron energy distribution. Errors given are one standard deviation from random sampling.}
\end{deluxetable} 

The jet parameters include Doppler factor ($\delta$), magnetic field strength ($B$), electron energy density ($w_{p_{soll},}$) minimum and maximum electron Lorentz factors ($\gamma_{min}$ and $\gamma_{max}$), and power-law index of the electron energy distribution ($n$). The Doppler factor, electron energy density, and minimum Lorentz factor collectively affect the predicted fluxes of all three spectral components, but the maximum Lorentz factor has little influence on the predicted fluxes. The magnetic field also affects the predicted fluxes of the synchrotron and SSC components. The minimum and maximum Lorentz factors determine the range of frequency values spanned by all three spectral components.

We constructed best-fit SEDs for all knots detected in X-rays, which resulted in three SEDs for PKS 1045$-$188, one for 8C 1849+670, and three for PKS 2216$-$038. Core SEDs were not created because of optical saturation and contamination. In all cases, the SSC component is several orders of magnitude too low to fit the knot fluxes, but the synchrotron and inverse-Compton components provide a reasonable fit for the radio and x-ray components, respectively. The best fit model parameters are not well constrained, as different combinations of parameters could achieve the same fits. With additional observations, the spectral index could be precisely determined, and with observations at the highest and lowest wavelengths of the distributions, the electron Lorentz factors could be better constrained. Unfortunately there are very few instruments with the requisite sensitivity and angular resolution capability that can obtain such data; therefore, most jet SED studies such as ours are currently limited in this way.

We estimated uncertainties in the fit parameters factors through sampling. We tested randomly chosen sets of parameters until we found 100 sets that successfully fit the observed data, and then we calculated the standard deviation for each parameter. Reported uncertainties are one standard deviation.

\section{\label{discussion}Discussion}

Although the SEDs do not uniquely determine the physical jet parameters, the low optical fluxes rule out the pure synchrotron emission for radio through X-ray. For PKS 1045$-$188 knots A and B, the optical upper limits are just high enough to allow a pure synchrotron model, but we stress that the majority of the fluxes of those upper limits likely comes from external contamination (Section \ref{optical}). We find that the X-ray emission from the jets of all the three blazars is consistent with the IC/CMB mechanism, but due to the small number of spectral points other X-ray emission models that do not contribute to the optical emission may work as well. In all three sources, the X-rays are found on the side with the approaching radio jet. The X-rays are detected on the FR\,I side (i.e., the side without a terminal hot spot) in 8C 1849+670 and PKS 2216$-$038, but on the FR\,II side (the side with a radio hot spot, e.g. knot C) in PKS 1045$-$188.

Although there are no detectable differences in the diffuse X-ray environment (Section \ref{xray}), for PKS 1045$-$188 and 8C 1849+670 in both cases the X-ray jets terminate significantly earlier than their corresponding radio jets (Fig. \ref{fig:1045rad} and \ref{fig:1849rad}). In addition, for PKS 1045$-$188, the radio jet bends significantly at the X-ray terminal point. The X-ray terminal peaks may indicate the sites of bulk jet deceleration, where the jet Doppler factor and the IC/CMB emission both drop dramatically.

We searched the literature for hybrid/asymmetric radio sources with Chandra data and found discording results as far as the X-ray emission mechanism was concerned (Table \ref{table:summary}). In only one case, 3C\,371 \citep{Sambruna07}, the X-ray emission was detected on the FR\,II side and was consistent with synchrotron emission. In all other cases, the X-ray emission was detected on the FR\,I side. It was consistent with synchrotron emission in 0521$-$365 \citep{Birkinshaw02}, 2201+044 \citep{Sambruna07}, NGC\,6251 \citep[especially the inner jet;][]{Evans05}, and it was consistent with IC/CMB in PG 1004+130 \citep{Miller06}, 3C\,433 and 4C\,65.15 \citep{Miller09}, 2007+777 \citep{Sambruna08}, and 3C\,17 \citep{Massaro09}. It is not clear if the X-ray emission is synchrotron or IC/CMB in the FR\,I jet of 1510$-$089 \citep{Sambruna04}. While the X-ray jet in this source is much shorter (less than half in extent) than the radio jet, similar to synchrotron jets in FR\,I sources, the lack of a clear optical counterpart favours the IC/CMB mechanism. 

\begin{deluxetable}{lccc}
\tablewidth{0pt}  
\tablecaption{Summary of Hybrid Source X-ray Jets\label{table:summary}}  
\tablehead{\colhead{Name} & \colhead{X-ray Jet Type} & \colhead{Emission Mechanism} & \colhead{Reference}}
\startdata
0521$-$365     & FR\,I  & Synchrotron & \citet{Birkinshaw02} \\
2201+044       & FR\,I  & Synchrotron & \citet{Sambruna07}   \\
NGC 6251       & FR\,I  & Synchrotron & \citet{Evans05}      \\
PG 1004+130    & FR\,I  & IC/CMB      & \citet{Miller06}     \\
3C 433         & FR\,I  & IC/CMB      & \citet{Miller09}     \\
4C 65.15       & FR\,I  & IC/CMB      & \citet{Miller09}     \\
2007+777       & FR\,I  & IC/CMB      & \citet{Sambruna08}   \\
3C 17          & FR\,I  & IC/CMB      & \citet{Massaro09}    \\
1510$-$089     & FR\,I  & IC/CMB      & \citet{Sambruna04}   \\
8C 1849+670    & FR\,I  & IC/CMB      & This paper           \\
PKS 2216$-$038 & FR\,I  & IC/CMB      & This paper           \\
3C 371         & FR\,II & Synchrotron & \citet{Sambruna07}   \\
PKS 1045$-$188 & FR\,II & IC/CMB      & This paper           \\
\enddata
\end{deluxetable} 

In 11 of the 13 studied hybrid sources, the FR\,I jet is the approaching, X-ray emitting jet. This trend could be a coincidence due to the small-number statistics, but if holds true after studies of additional hybrid sources then it could seriously question theories regarding the creation and the true nature of hybrid sources. Those theories that rely on asymmetries, such as asymmetric environments \citep{Miller09} or asymmetric magnetic fields in the accretion disk \citep{Wang92}, would not be able to explain why the FR\,I jet would be the approaching jet in the majority of cases. Other theories would remain, such as if hybrid sources were actually relativistic FR\,II sources with bent approaching jets, then the inner portion of the jet could be beamed toward the observer and the terminal hotspot could be beamed in a different direction, giving the approaching jet the appearance of an "edge-darkened" FR\,I. Again, we must emphasize that until statistics are improved, nothing can be conclusively ruled out, so studies of additional hybrid sources are vital.

\begin{figure}
\centering{
\includegraphics[width=16cm]{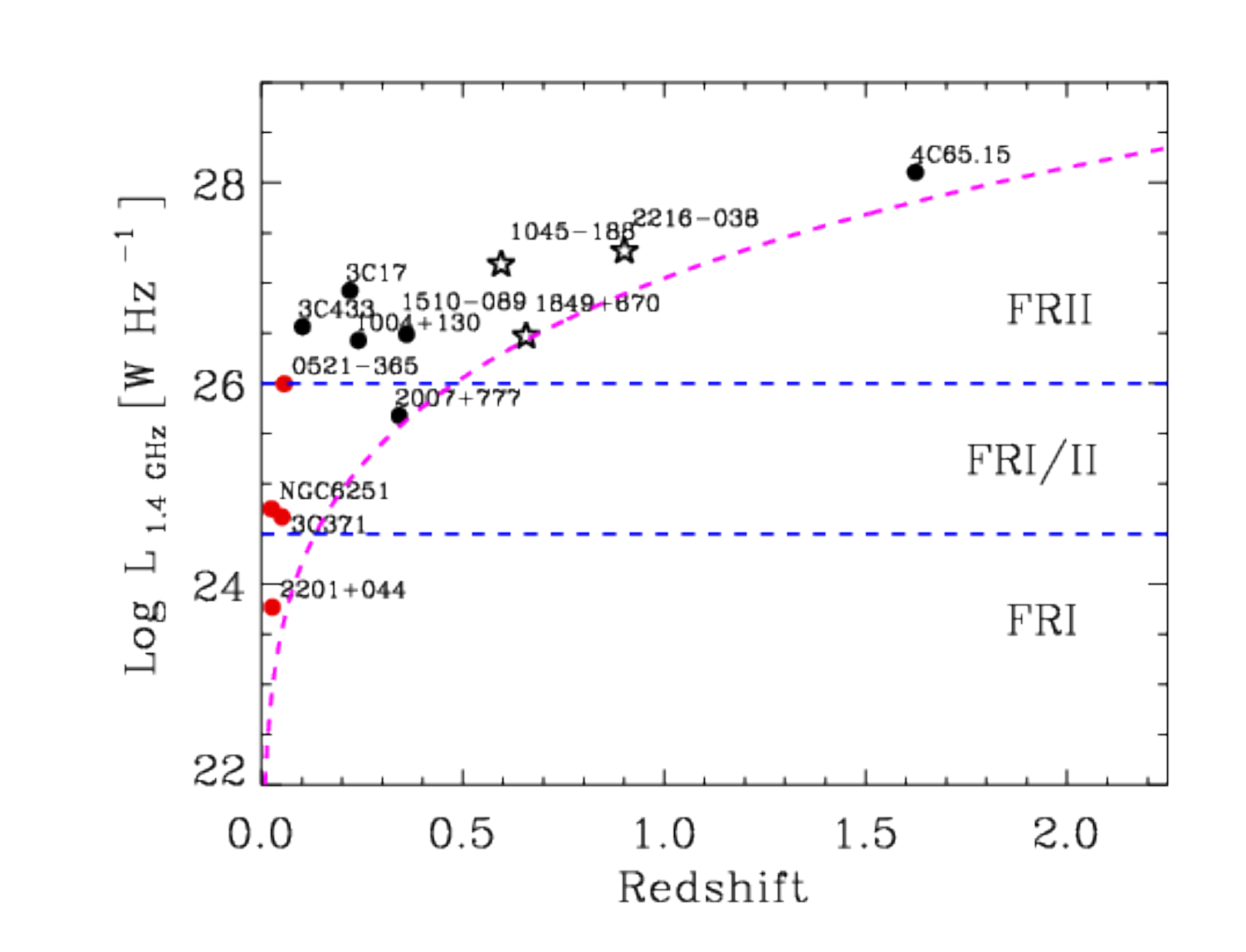}}
\caption{Total radio power at 1.4~GHz plotted against redshift for all hybrid jet AGNs in the literature. The red symbols denote sources with X-ray jets consistent with the synchrotron mechanism, and the black symbols denote those consistent with IC/CMB. The open stars denote the three sources presented in this paper. The dashed lines demarcate FR classification by \citet{Ledlow96}. The magenta line indicates the detection limit of the VLSS.}
\label{fig:power}
\end{figure}

Our observations indicate that even jets with low power FR\,I-like appearance have X-rays from IC/CMB when the total radio power of the source is like an FR\,II. This could imply that even on the FR\,I side there is a fast collimated inner spine, just like an FR\,II jet, but which dissipates or is de-beamed before a terminal hot spot is formed. It appears that, rather than the FR morphology, the total radio power is the main determinant for the X-rays being IC/CMB (in high power sources) or synchrotron (in low power sources). This is supported by the plot of total radio power versus redshift (Figure~\ref{fig:power}). With the exception of 2007+777 (which is a borderline FR\,I/II source), all sources with FR\,II-like total radio powers at 1.4~GHz have X-ray emission from the IC/CMB mechanism. In the remaining sources, synchrotron emission dominates. Furthermore, in all of the hybrid sources studied, regardless of emission mechanism, the X-ray emission is from the approaching jet. This is in good agreement with previous studies, which indicate that blazars are excellent candidates for X-ray jet detections \citep{Hogan11}.

\section{\label{conclusion}Conclusions}
In this paper we present  multiwavelength jet observations of PKS 1045$-$188, 8C 1849+670, and PKS 2216$-$038, three radio-loud AGN from the MOJAVE-Chandra Sample that straddle the Fanaroff-Riley boundary. These hybrid sources are of interest because they provide an excellent opportunity to study the jet emission mechanisms and the influence of the external environment.

We used archival VLA observations at 1.4, 4.6, and 8.4 GHz; deep HST observations with the F160W and F475W filters; and deep Chandra observations. We used MOJAVE VLBA parsec-scale observations to identify the approaching jets and VLA kiloparsec-scale observations to classify the approaching jets as FR\,II for PKS 1045$-$188 and FR\,I for 8C 1849+670 and PKS 2216$-$038. From the Chandra observations, we identified X-ray emission from three of five radio-visible knots in PKS 1045$-$188, one of two knots in 8C 1849+670, and all three knots in PKS 2216$-$038. We also identified X-ray emission from the area of the counter jet of PKS 1045$-$188, but it is too faint and diffuse to identify any specific emission regions without further observation.

For the seven X-ray visible knots, we constructed and fit SEDs using the synchrotron and IC/CMB emission models of \citet{Krawczynski04}. Although we would require observations at additional wavelengths to fully constrain the model parameters, we found that the weak optical emission ruled out synchrotron emission for radio to X-ray in all cases, even for the FR\,I jets of 8C 1849+670 and PKS 2216$-$038.

All three sources have high total extended radio power, similar to that of FR\,II sources. We find this is in good agreement with previously studied hybrid sources, where high-power hybrid sources emit X-rays via IC/CMB and the low-power hybrid sources emit X-rays via synchrotron emission. This supports the idea that it is total radio power rather than FR morphology that determines the X-ray emission mechanism. Additionally, the X-ray-emitting jet is the approaching jet in all three sources, which is also in good agreement with previously studied hybrid sources.

We found no significant asymmetries in the X-ray environments. Sources PKS 1045$-$188 and 8C 1849+670 show significant differences in their radio and X-ray termination points which may be locations of bulk deceleration.

Questions remain regarding why the FR\,I jet is the approaching, X-ray-emitting jet in 11 of the 13 studied hybrid sources. This could still be coincidence due to the small number of studied hybrid sources, or this could indicate additional effects, such as the possibility that some of the FR\,I jets are actually bent, de-beamed FR\,II jets. Observations of additional sources are necessary to answer these questions.

\acknowledgments

We thank the anonymous referee for a careful review that improved this paper. This work was supported by the National Aeronautics and Space Administration (NASA) through Chandra Award Numbers (GO3-14120A, GO3-14120B) issued by the Chandra X-ray Observatory Center (CXC), which is operated by the Smithsonian Astrophysical Observatory (SAO) for and on behalf of NASA under contract NAS8-03060. Support for program number 13116 was provided by NASA through a grant from the Space Telescope Science Institute, which is operated by the Association of Universities for Research in Astronomy, Inc., under NASA contract NAS5-26555. The MOJAVE program is supported under NASA-Fermi grant NNX12A087G. The National Radio Astronomy Observatory is a facility of the National Science Foundation operated under cooperative agreement by Associated Universities, Inc.

{\it Facilities:} \facility{VLA}, \facility{HST (WFC3)}, \facility{CXO (ACIS)}.


\bibliographystyle{apj}
\bibliography{hybrid_blazars}

\clearpage

\end{document}